\documentclass[fleqn,12pt]{elsarticle}

\usepackage{graphicx,color}
\usepackage{amsmath,amssymb}
\newcommand{\Slash}[1]{{\ooalign{\hfil/\hfil\crcr$#1$}}}

\newcommand{\Tr}{{\rm Tr}}
\newcommand{\Ln}{{\rm Ln}}

\newcommand{\lqcd}{\Lambda_{\rm QCD}}

\newcommand \beq{\begin{equation}}
\newcommand \eeq{\end{equation}}
\newcommand{\vp}{\vec{p}}

\newcommand{\para}{\parallel}

\newcommand{\calG}{\mathcal{G}}

\newcommand{\calV}{\mathcal{V}}

\newcommand{\rmd}{\mathrm{d}}
\newcommand{\rmi}{\mathrm{i}}
\newcommand{\rme}{\mathrm{e}}

\newcommand{\tp}{ \tilde{p} }

\journal{Physics Letter B}

\begin{document}

\begin{frontmatter}
\title{Renormalizing the zero point energy in dense QCD}
\author {Toru Kojo}
\address {Key Laboratory of Quark and Lepton Physics (MOE) and Institute of Particle Physics, Central China Normal University, Wuhan 430079, China}
%
\begin{abstract}
We analyze the zero point energy in a dense matter of quarks or hadrons with particular attention on the renormalization of the UV divergences. Besides divergences removable by the vacuum subtraction and counter terms, there are also UV divergences associated with non-perturbative modifications of quark bases appearing in the in-medium propagators. The latter would remain after the self-energies and vertices are renormalized, unless a proper set of medium contributions is included at a given truncation. We use the formalism of the two particle irreducible action to clarify how the UV divergences are assembled to cancel. An example is given for the thermodynamic potentials with mesons as composite particles whose zero point energies apparently diverge but can be cancelled by the quark self-energy contributions. Important applications of this work are quark matter with hadronic correlations which may be realized at the core of neutron stars.

\end{abstract}
\end{frontmatter}

\section{Introduction}
\label{sec:intro}

The equation of state of cold, dense matter in quantum chromodynamics (QCD) is a crucial ingredient to understand astrophysical objects such as neutron stars \cite{Baym:2017whm}. The equation of state generally contains the zero point energy, a sum of energies over all possible vacuum fluctuations. In spatially three dimensions, the zero point energy in a matter has the quartic UV divergence $\sim \Lambda_{{ \rm UV}}^4$ ($\Lambda_{{ \rm UV}}$: UV cutoff) which must be canceled by subtracting out the zero point energy in vacuum. Since the pressure relevant for phenomenological applications (e.g. neutron stars)  is at most $\sim 1\, {\rm GeV/fm^3}$ (which is governed by a quark chemical potential $\mu$ and the QCD non-perturbative scale $\lqcd$), the UV divergences should be handled in a complete manner, otherwise the big number would totally contaminate our predictions.

The cancellation of the UV divergences is intricate. The leading contributions are given by the particles at large energy which are insensitive to the properties of a medium. Thus the resulting quartic divergences are universal and can be cancelled by a simple vacuum subtraction. But there would remain the quadratic and logarithmic divergences with the coefficients dependent on a medium. Naive power counting suggests quadratic divergences, e.g.,
\beq
\sim  \left(M^2_{{\rm med}} - M^2_{{\rm vac}} \right) \Lambda^2,~~~~~ \sim \mu^2 \Lambda^2 \,,
\eeq
where the first one is associated with the difference in the medium and vacuum effective masses, while the second solely depends on the chemical potential. 

In perturbative calculations with Feynman diagrams \cite{Freedman:1976xs,Kurkela:2009gj}, only the UV divergences of the latter type must be handled. As known from textbook examples \cite{Kapusta}, for {\it each} Feynman graph the divergences in sub-graphs, such as in the self-energies and vertices, can be made finite by the vacuum counter terms and the rest of the divergences can be subtracted off by the vacuum graph with the same topology. Then the resulting contributions are such that the energy of a quark line is essentially bound by $\mu$. Schematically, with the vacuum subtraction we encounter the following terms,
($\int_{\vp} = \int \rmd^3 \vp /(2\pi)^3$)
\beq
\int_{\vp} F(\vp) \left[ \theta(E (\vp) - \mu) - 1 \right] = \int_{\vp} F(\vp) \, \theta \left( \mu -E (\vp) \right)   \,,
\label{eq:subtract1}
\eeq
where the first and second terms in the first equation come from the medium and vacuum quark propagators respectively. With the {\it same} renormalized coefficient $F(\vp)$ of the mass dimension 1, they are combined to yield the UV finite contributions. It is very remarkable that the cancellations are much stronger than expected from the power counting of momenta; not only quartic but also quadratic and logarithmic divergences cancel. 

\begin{figure*}[!t]
\begin{center}
\vspace{-1.5cm}
\hspace{-.3cm}
\includegraphics[width = 0.5\textwidth]{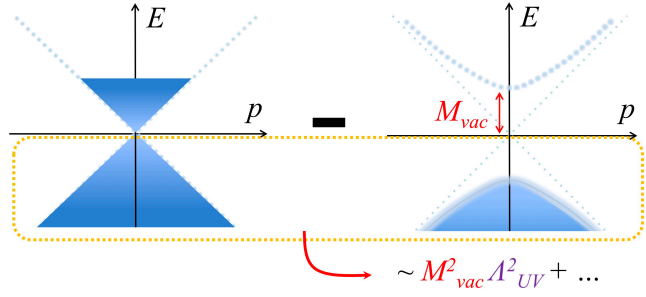}
\vspace{-0.4cm}
\end{center}
\vspace{-1.5cm}
\caption{The single particle contribution of quarks (or baryons) to the thermodynamic potential at high density (left) and in the vacuum (right). The difference in the Dirac sea results in the quadratic UV divergence $\sim \Lambda_{ {\rm UV} }^2 $. 
\footnotesize{
}
\vspace{-0.0cm} }\label{fig:Dirac_sea}
\end{figure*}

The above UV cancellations become incomplete, however, when different quark bases, e.g. with different residues and effective masses, are used for the vacuum and at finite density (Fig.\ref{fig:Dirac_sea}). Attaching $*$ to the medium modified quantities, Eq.(\ref{eq:subtract1}) changes into
\beq
\int_{\vp} \left[ F_* (\vp)\theta(E_* (\vp) - \mu) - F (\vp) \right] = \int_{\vp} F_* (\vp) \, \theta \left( \mu -E (\vp) \right) +  \int_{\vp} \left[ F_* (\vp) - F(\vp) \right]\,,
\eeq
where the last term is associated with the changes in bases. It is plausible to have the asymptotic behavior $F_*(\vp) \sim F(\vp) + c /\vp^2$ at large $|\vp|$, with the coefficient $c$ reflecting the difference of the bases. But even with this damping behavior there remain the quadratic UV divergences.

To handle this sort of UV divergences, it is necessary to keep track all changes generated by modifications of bases. In this paper we will argue how to assemble these UV divergences to cancel. In particular we discuss which contributions should be included together in a given truncated calculation. For example, in calculations to include vacuum fluctuations from both quarks and hadrons as composite particles, the UV divergences in the zero point energies of quarks can be cancelled with those from hadrons, if both contributions are treated in a consistent manner.

\section{The 2PI action}
\label{sec:2PI}

To keep track all impacts associated with modifications of quark bases, it is useful to use the formalism of the two particle irreducible (2PI) action, developed by Luttinger-Ward \cite{Luttinger:1960ua}, Baym-Kadanoff \cite{Baym:1961zz,Baym:1962sx}, and its relativistic version by Cornwall-Jackiw-Toumbolis \cite{Cornwall:1974vz}.  The renormalizability of the 2PI action was first discussed by van Hees and Knoll for scalar field theories at finite temperature \cite{vanHees:2001ik}, and the analyses were refined in seminal works \cite{Blaizot:2003an}. The analyses have been further extended to generate the resummed vertices consistent with the Ward identities \cite{Berges:2005hc}. The QED case was discussed in \cite{Reinosa:2009tc}. In the presence of chemical potential, the charged scalar field theories were discussed in \cite{Marko:2014hea} where the authors clarified how the Silver Blaze property is maintained in the 2PI formalism. 
On the other hand, to the best of our knowledge, the renormalizability of fermion theories with finite chemical potentials has not been discussed explicitly, and we will discuss it below. 

The 2PI action $I[S;\mu]$ is given as a functional of a quark propagator $S$,
\beq
I [S;\mu] =  \Tr \Ln S + \Tr\left[ S \left(S^{-1}- ( S^{\mu}_{ {\rm tree} } )^{-1} \right) \right] + \Phi[S] \,,
\eeq
which includes a tree level propagator (in Euclidean space), $S^{\mu}_{ {\rm tree} } (p) = S^{\mu=0}_{ {\rm tree} } (\tp) = -\left( \Slash{\tp} - m \right)^{-1}$, with the current quark mass $m$ and with momenta $\tp_\mu = (p_0-\rmi \mu, p_j)$.
 The $\Phi$ functional is the sum of 2PI graphs composed of dressed quark propagators\footnote{We can also consider the functional of gluon propagators, but to take the shortest path to illustrate our points we consider only $S$ as variables.} and the counter terms necessary for the renormalization of  the vacuum self-energies and vertices. The variation of $\Phi$ with respect to $S$ yields the graphs for the quark self-energies, and if we choose $S$ to give the extrema, it satisfies the Schwinger-Dyson equation,
\beq
\frac{\, \delta I [S;\mu] \,}{\, \delta S (p) \,} 
= S^{-1} (p)- (S^{\mu}_{ {\rm tree} } )^{-1}(p) - \Sigma^{[S]} (p) = 0 \,,~~~~~~~ \Sigma^{ [S] } (p) = - \frac{\, \delta \Phi [S] \,}{\, \delta S (p) \,} \,.
\eeq
The solution of the Schwinger-Dyson equation is written as $S^{\mu}_* (p)$ and the corresponding self-energy as $\Sigma^{\mu}_* (p) \equiv \Sigma^{ [S^{\mu}_*] } (p) $. If we further differentiate $\Phi[S]$ by $S$, we get kernels with the 3-, 4-, and more-external legs.  

As far as we concern functionals of propagators but not vertices, the renormalization programs of the vacuum self-energies and vertices are the same as the BPHZ scheme \cite{BPHZ,BPHZ2}, except replacement of perturbative propagators with the full ones. Strictly speaking the full propagators should be such that their asymptotic behaviors do not spoil the power counting used for the BPHZ, and we will consider only such class of solutions for $S^{\mu=0}_*$.
Below we assume that these renormalizations were already done in the vacuum so that $S^{\mu=0}_*, \Sigma^{\mu=0}_* , \cdots$ are all renormalized, and the counter terms are all fixed. Then no additional counter terms are necessary for in-medium self-energies and vertices; for instance in solving the Schwinger-Dyson equation for the in-medium self-energy, one can expand $\Sigma^{[S]}$ around $S^{\mu=0}_*$,
\beq
\Sigma^{[S ]} (p) = \Sigma^{[S^{\mu=0}_* ]} (p) + \int_k \frac{\, \delta \Sigma^{ [S^{\mu=0}_*] } (p) \,}{\, \delta S (k) \,} \left( S (k) -S^{\mu=0}_* (k) \right) + \cdots \,,
\eeq
where the logarithmic divergences as well as the counter terms are hidden in the vacuum quantities. The first term in the RHS is the renormalized vacuum self-energy and is finite. The second term is a renormalized 4-point vacuum function whose two legs are contracted with $S-S^{\mu=0}_*$. Considering the fact that the contraction of $S$ with two legs of 4-point function yields only the logarithmic divergence, the replacement $S \rightarrow S-S^{\mu=0}_*$ is enough to get the UV finite result, as far as we consider a class of solutions such that $S-S^{\mu=0}_* \sim 1/p^2$ at large momenta. The other $n$-point functions will be treated in the same say.

Now we consider our main concern in this paper, the $0$-point function,
\beq
I_R [S;\mu] \equiv I[S ;\mu] - I[S^{\mu=0}_* ; \mu=0] \,.
\eeq
In order to show that the vacuum subtraction makes the $0$-point function UV finite, we first make a decomposition,
\beq
I_R [S;\mu] = \left( I[S ;\mu] - I[S^\mu_\parallel ;\mu] \right) - \left( I[S^\mu_\parallel ;\mu] - I[S^{\mu=0}_* ; \mu=0] \right)
\equiv I_{\Delta S} [S;\mu] + I_{\Delta \mu}\,.
\eeq
where we have introduced an in-medium propagator made of quark bases optimized for the vacuum,
\beq
 S^\mu_{\parallel} (p) \equiv S^{\mu=0}_* ( p \rightarrow \tp) = \left[ - \Slash{\tp} + m +  \Sigma^{\mu}_{\parallel} (p) \right]^{-1} \,,~~~~~~  \Sigma^{\mu}_{\parallel} (p) \equiv \Sigma^{\mu=0}_* (p \rightarrow \tp) \,.
\eeq
We note that in general $S^\mu_\parallel$ does not minimize $I[S;\mu]$ hence $\Sigma^{\mu}_\parallel$ differs from $-\delta \Phi/\delta S |_{ S_{\para}^\mu }$. 

\begin{figure*}[!t]
\begin{center}
\vspace{-1.5cm}
\hspace{-.3cm}
\includegraphics[width = 0.3\textwidth]{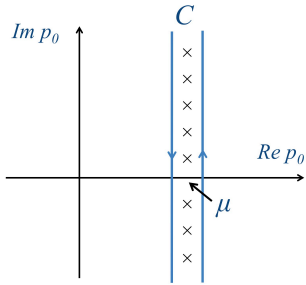}
\vspace{-0.cm}
\end{center}
\vspace{-0.5cm}
\caption{ 
\footnotesize{The contour $C$ taken in Eq.(\ref{Eq:contour}).
}
\vspace{-0.0cm} }
\label{fig:contour}
\end{figure*}

The functional $I_{\Delta S} [S;\mu]$ is the energy cost or gain associated with the change of bases from the vacuum one. Clearly at $S=S_\parallel^\mu$ the functional is  $I_{\Delta S} =0$, showing that at least for some class of $S$ the functional $I_{\Delta S}$ is UV finite. So the rest of the question is whether $I_{\Delta \mu}$ is UV finite. This can be discussed following a conventional treatment. For the moment we introduce an infinitesimal temperature $T$ and use the Matsubara formalism with $\tp \rightarrow \omega_n - \rmi \mu$ where $\omega_n = (2n+1)\pi T$. Noting that the $\mu$ and $p_0$ always appear together in the combination $\tp_0$, one can isolate the vacuum piece in $I[S_\para^\mu;\mu]$ by writing it in the form of $I = \Tr F$, 
\beq
I[S_\para^\mu;\mu] = T \sum_n \int_{\vp} F_{\vp} ( \rmi \omega_n + \mu)
= \int_C \frac{\, \rmd p_0 \,}{ 2\pi \rmi } \frac{F_{\vp}(p_0 ) }{\, \rme^{\beta (p_0 -\mu) }+ 1 \,}  
\rightarrow \int_{\vp} \sum_{ E_{\vp} } \left[ 1 + \theta(\mu-E_{\vp} ) \right] F_{\vp} (E_{\vp})
\,,
\label{Eq:contour}
\eeq
where the contour $C$ is shown in Fig.\ref{fig:contour} and $E_{\vp}$'s are possible poles whose values are bound by $\mu$. Here we are using the quark bases same as the vacuum, so the function $F_{\vp}$ takes the exactly same form as in the vacuum, hence the UV divergent piece, which is not bound by $\mu$, can be completely cancelled by the vacuum subtraction. With this we conclude the function $I_{\Delta \mu}$ is UV finite.

\section{The structure of $I_{\Delta S}$}
\label{sec:structure}

Our main concern in this paper is the UV contributions associated with the change of bases, which are non-perturbative effects summarized in the functional $ I_{\Delta S} [S;\mu]$. By construction $I_{\Delta S} [S;\mu]$ is guaranteed to be UV finite, as $I_{\Delta S}$ can be made zero by choosing $S=S^\mu_\parallel$. As $S$ departs from $S^\mu_\parallel$, the UV contributions come into play. The question is how much these propagators can be different, or what class of $S$ is acceptable. In principle by solving the Schwinger-Dyson equation one automatically rejects $S$ which yields the UV divergence. Nevertheless the consideration of this question is important in practice. For example, in numerically solving the Schwinger-Dyson equations by iterative methods, the understanding of the UV behavior helps us to find efficient trial functions to begin with. Or for some $\Phi$-functional even such efficient Ansatz costs too much time; in that case we need to restrict the possible form of $S$ by introducing finite number of variational parameters, or by relying on some physical insights. For the description of hadronic correlations in quark matter, the $\Phi$-functional we will need  should have infinite number of loops, so approximations for $S$ should be as simple as possible.

For these reasons we analyze the UV structure of the functional $I_{\Delta S}$ and the condition required for $S$. Let us first focus on the ``single particle" contribution $\Tr \Ln S$, for which
\beq
\Tr \Ln S - \Tr \Ln S^\mu_\parallel  = - \Tr \Ln \left(1 + S^\mu_\parallel \left(S^{-1}- (S_\para^{\mu})^{-1} \right) \right) 
\sim - \Tr \left[ S^\mu_\parallel \left( S^{-1} - (S^\mu_\parallel)^{-1}  \right) \right] \,,
\label{eq:single}
\eeq
whose contribution diverges unless the difference of the self-energies of $S$ and $S^\mu_\para$ damps faster than the power of $\sim \Slash{p}/p^4$, with the extra $1/p^4$ suppression. But this requirement looks too strong: although the infrared effects likely decouple at high energy, from the dimensional ground it is natural to expect the extra suppression of $\sim 1/p^2$. For instance when the BCS type gap appears by medium effects\footnote{The momentum independence of the BCS gap is the consequence of contact interactions. If we use more realistic long-range interactions the gap functions decrease at large momenta.}, the propagator for a quasi-particle is $ p_0 -\rmi \sqrt{ (E -\mu)^2 + \Delta^2 } \sim \tp_0 - \rmi (E + \Delta^2/|\vp|)  $ at large momenta, and the gap effects damp as $\sim \Slash{p}/p^2$, which is too slow to make Eq.(\ref{eq:single}) UV finite. 

This level of strong damping is also required for the terms in $I_{\Delta S}$ other than $\Tr \Ln S$ to be free from the UV divergences.
Actually, the apparent strong UV sensitivity is eliminated by assembling these contributions, and it turns out that $I_{\Delta S}$ can be expressed by powers of $S-S^\mu_\parallel$ starting with the quadratic order, as we will check below. This relaxes the condition on the solutions $S=S_*$, and makes the physics in the infrared much less dependent on the UV physics. 

For the sake of arguments  we will count the degree of divergence for a class of $S$ whose asymptotic behavior is $S^{-1} - (S^\mu_\parallel)^{-1} \sim \Slash{p} \Delta^2 /p^2$. Here $\Delta$ is some non-perturbative scale in the infrared.
We first isolate the leading UV contribution of $\sim \Delta^2 \Lambda_{ {\rm UV} }^2$ from the ``single particle" contribution, 
\beq
\Tr \left( \Ln S - \Ln S^\mu_\parallel \right)
= I_{L_1} [S] - \Tr \left[ S^\mu_\parallel \left( S^{-1} - (S^\mu_\parallel)^{-1}  \right) \right] \,,
\eeq
where 
\beq
I_{1} [S] \equiv \Tr \left( \Ln S - \Ln S^\mu_\parallel \right) + \Tr \left[ S^\mu_\parallel \left( S^{-1} - (S^\mu_\parallel)^{-1}  \right) \right]
~\sim~  \Tr \left[ S^\mu_\parallel \left( S^{-1} - (S^\mu_\parallel)^{-1}  \right) \right]^2 \,.
\eeq
For $S$ being discussed the functional $I_{1}$ behaves as $\sim \Delta^4 \ln \Lambda_{ {\rm UV}}$. Next, the isolated piece which would have quadratic divergence is combined with the other terms in $I_{\Delta S}$, leaving us with
\begin{align}
I_2 [S]  & \equiv - \Tr \left[ S^\mu_\parallel S^{-1} - 1 \right]
 +  \Tr\left[ 1 - S (S^{\mu}_{ {\rm tree} })^{-1} \right] 
 -  \Tr\left[ 1 - S^\mu_\parallel ( S^{\mu}_{ {\rm tree} })^{-1}  \right] 
+ \Phi[S] - \Phi[S_{ \para }^{\mu}] 
\nonumber \\
& = \Phi[S] - \Phi[S_{ \para }^{\mu}] + \Tr\big[ \left(S-S_{ \para }^{\mu} \right) \left( S^{-1} - (S^{\mu}_{ {\rm tree} } )^{-1} \right) \big] \,.
\end{align}
Expanding $\Phi[S]$ around $S^\mu_\para$, 
\beq
I_2[S]
= \Tr\left[ \left(S-S_{ \para }^{\mu} \right) \left( S^{-1} - (S^{\mu}_{ {\rm tree} } )^{-1} - \Sigma^{ [S_\para^\mu ] } \right) \right] 
+ \sum_{n=2} \frac{\, 1 \,}{ n!} \Tr\left[ \left(S - S_{ \para }^{\mu} \right)^n  \frac{\, \delta^n \Phi \,}{\, \delta S^n \,} \bigg|_{S_{ \para }^{\mu} } \right]
\,.
\eeq
(Reminder: in general $\Sigma^\mu_\para \neq \Sigma^{ [S^\mu_\para] }$ so that $S_\para^\mu \neq ( S_{\rm tree}^{\mu ~-1} + \Sigma^{ [S_\para^\mu] } )^{-1}$.)
The first term would be quadratically divergent. But for a class of $S$ that approaches $ (S^{\mu}_{ {\rm tree} } )^{-1} + \Sigma^{ [S_\para^\mu ] }$ at large momenta, the degree of divergence reduces to the logarithmic one. For solutions of the Schwinger-Dyson equation, $S^\mu_*$, we have
\beq
\Sigma^{[S^\mu_*] }- \Sigma^{[S^\mu_\para] }= - (S^\mu_* - S^\mu_\para) \frac{\, \delta^2 \Phi \,}{\, \delta S^2 \,} \bigg|_{S_{ \para }^{\mu} } + \cdots \,,
\eeq
so $I_2[S^\mu_*]$ start with the quadratic powers of $S_*-S_\para$, hence is the order of $ \sim \Delta^4 \ln \Lambda_{ {\rm UV}}$.

Thus we have verified that, if $S^{-1} - (S^\mu_\parallel)^{-1} \sim \Slash{p} \Delta^2 /p^2$, the energy shift $I_{\Delta S} = I_1 + I_2$ associated with the changes in bases is the order of $\sim \Delta^4 \ln \Lambda_{ {\rm UV}}$. The rest of our task is to examine the possible logarithmic divergences. Although we have looked for the possibility of their cancellations, we were not able to find general reasoning why this is the case. So at this point we are forced to think of the following possibilities: (i) the infrared effects decouple faster than the behavior $S^{-1} - (S^\mu_\parallel)^{-1} \sim \Slash{p} \Delta^2 /p^2$; (ii) the non-perturbative effects in the infrared decouple as expected in naive power counting, but each of $S^\mu_*$ and $S^\mu_\para$ has a damping behavior faster than $\sim 1/\Slash{p}$, that is, has the positive anomalous dimension as found in the perturbative results improved by the renormalization group. We also note that in asymptotic free theories the kernel, $\delta^n \Phi/\delta S^n$, approaches the weak coupling results in which the coupling constant zero at high energy. This also tempers the logarithmic divergences. While these possibilities seem to us feasible in asymptotic free theories, the detailed discussion on these effects is beyond the scope in this paper.


%

\section{The zero point energy of composite particles}
\label{sec:composite}

\begin{figure*}[!t]
\begin{center}
\vspace{-1.5cm}
\hspace{-.3cm}
\includegraphics[width = 0.5\textwidth]{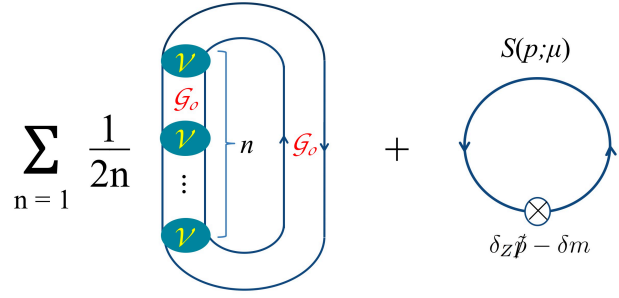}
\vspace{-0.4cm}
\end{center}
\vspace{-1.5cm}
\caption{. 
\footnotesize{The resummed diagram producing mesonic states and the counter term contracted with a full propagator.
}
\vspace{-0.0cm} }\label{fig:2PI_hadrons}
\end{figure*}

While the preceding sections on the 2PI action may not be quite new for experts familiar with this formalism, the cancellation of quadratic divergences between single particle contributions and the rest is hardly trivial from qualitative point of view. This can look more spectacular in discussions of the 2-, 3-, or more particle correlations with resonance poles \cite{Dashen:1969ep,Blaschke:2013zaa} which arise from the resummation of graphs  in the $\Phi$-functional. If we pick up one of mesonic states in vacuum and medium, their energy difference contributes the quadratic divergence to the energy,
\beq
\sim \int_{\vp} \left( \sqrt{ m_*^2 + \vp^2 } - \sqrt{ m_{ {\rm vac} }^2 + \vp^2 } \right) \sim \left(m_*^2 - m_{ {\rm vac} }^2 \right) \Lambda_{ {\rm UV} }^2 \,.
\label{eq:meson}
\eeq
(This belongs to the power counting assumed in the previous section.)
The situation becomes even more complicated when we start to take into account the disappearance of hadronic states and their decays into quarks. It looks very difficult to handle these UV effects starting with hadronic descriptions then putting the quark substructure effects as corrections, because there are too much freedom or no guideline in purely hadronic descriptions. Rather it is safer to start with quark descriptions to write down the energy functional insensitive to the UV physics, and then focus on some relevant hadronic contributions generated in the $\Phi$-functional. In this respect the 2PI action has the advantage, as the functional $I_1$ and $I_2$ could be written in the form where the most dangerous quadratic UV divergences is already cancelled, allowing us to focus on the physical effects in the infrared.

To be concrete we take up 2-body correlations as an example. The corresponding $\Phi$-functional (with the counter terms) is shown in Fig.\ref{fig:2PI_hadrons}. This kind of resummation has been standard in the context of the BEC-BCS crossover \cite{NSR,Diener,Ohashi:2002zz} in which it is important to construct composite bosons from fermions. Often used approximation for the thermodynamic potential includes the contributions from the mean field single particle energy plus the Gaussian pair fluctuations. Its relativistic extension was attempted in \cite{Abuki:2006dv,Sun:2007fc}. From the viewpoint of the 2PI framework, however, it does not treat the quark self-energy and $\Phi$-functional consistently and the analyses would suffer from the quadratic divergences. In non-relativistic theories the problems of the quadratic divergences do not appear because the fermion propagator damps as $\sim 1/\vp^2$. In relativistic theories, however, one must go beyond this approximation; it is necessary to use the fermion self-energy including pair fluctuations which is used to cancel the quadratic divergences in the $\Phi$-functional, e.g. Eq.(\ref{eq:meson}).

The 2PI graphs for mesons include 2-quark lines (Fig.\ref{fig:2PI_hadrons}). The free streaming 2-body propagator is given by
\beq
\calG^0_{\alpha\alpha' ; \beta \beta'} ( p_+, p_- ) = S_{\alpha \alpha'} ( p_+) S_{\beta \beta'} ( p_- ) \,, ~~~~~~~~ p_\pm = p \pm \frac{\, k \,}{2} \,,
\eeq
where $k$ is the total momentum of a meson which is conserved, while $p$ is the relative momentum. Using the 2PI vertices with 4-legs, the resummed 2-body propagator is given in a symbolic form as
\beq
\calG = \calG_{0} +  \calG_{0} \calV \calG_{0} + \cdots = \frac{ \calG_0 }{\, 1 - \calV \calG_0 \,} \,,
\eeq
where the summation over internal momenta and other quantum numbers is implicit. Meanwhile, in the 2PI diagrams we must attach the symmetry factor $1/(2n)$ for a diagram with $n$-2PI vertices, so (the trace runs over the momenta $p$ and $k$ and other quantum numbers)
\beq
 \Phi_{ {\rm M} } [S] = \sum_{n=1} \frac{1}{\, 2n \,} \Tr\left[  \left( \calV \calG_0 \right)^n \right]
 = - \frac{1}{\, 2 \,} \Tr \left[ \Ln \left( 1- \calV \calG_0 \right) \right] 
 = \frac{1}{\, 2 \,} \Tr\Ln  \left( \calG/\calG_0 \right)^{-1} \,.
\eeq
The function $\calG/\calG_0 $ may be identified as a mesonic propagator sandwiched by effective quark-meson vertices. One also has the functional for the counter terms
\beq
 \Phi_{{\rm counter} } [S] = \Tr \left[ \left( \delta_Z  \Slash{ \tp} - \delta m \right) S \right] \,,
\eeq
which removes the logarithmic UV divergence in the Yukawa theory type diagrams (one loop diagram made of a mesonic line and a quark line).

\begin{figure*}[!t]
\begin{center}
\vspace{-1.5cm}
\hspace{-.3cm}
\includegraphics[width = 0.45\textwidth]{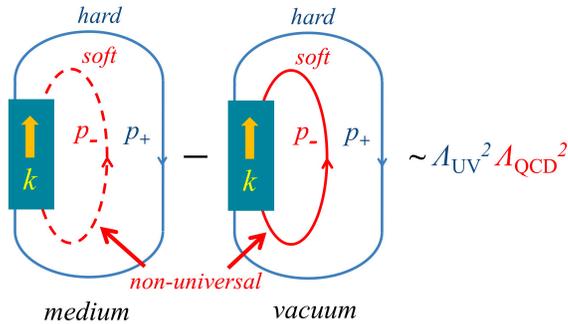}
\vspace{-0.cm}
\end{center}
\vspace{-1.5cm}
\caption{ 
\footnotesize{
The quadratic divergence in mesonic loop (made of 2-quark lines with the total momentum $k$) that survives after the vacuum subtraction. The hard component (with $p_+ \gg \lqcd,\mu$), which is universal at finite $\mu$ and in the vacuum, couples to the non-universal soft components (with $p_- \sim \lqcd,\mu$), leaving the medium dependent divergences.
}
\vspace{-0.0cm} }\label{fig:hadron_loop}
\end{figure*}

As advertised, there remain quadratic divergences after the vacuum subtraction. One might think that realistically the quark self-energies inside of the loops damp at large momenta and it would temper the divergences, but in fact those properties are not sufficient; this can be seen (Fig.\ref{fig:hadron_loop}) by considering the domain of the integration such that $p_+ \rightarrow \infty$ but $p_- \sim \lqcd$ in which the universal part of the quark propagator couples to the non-universal infrared part. This leaves us with the UV divergences with the coefficients dependent on a medium.

These are the quadratic divergences cancelled in the functional $I_2[S]$. The self-energy is
\beq
 \Sigma^{ [S] } (p) = - \frac{\, \delta \Phi_{ {\rm M} + {\rm counter} } [S] \,}{\, \delta S (p) \,} 
 = \frac{1}{\, 2 \,}  \left[ \frac{1}{\, 1- \calV \calG_0 \,} \frac{\, \delta \left( \calV \calG_0 \right) \,}{\, \delta S (p) \,} \right] - \left( \delta_Z  \Slash{ \tp} - \delta m \right) \,.
\eeq
In the functional $I_2[S]$, this term is contracted with $(S-S_\para)$ and then is subtracted from $\Phi[S]$. The subdiagram which becomes non-universal when soft momenta flow into is taken care by the part $S-S_\para$, and as a result the quadratic UV divergences shown in Fig.\ref{fig:hadron_loop} are cancelled in the functional $I_2$.
 
\section{Summary}
\label{sec:summary}

We have discussed how to perform UV finite computations at finite quark density. In particular we analyze how to cancel the would-be UV divergences associated with the changes in bases. The expression is grouped into several distinct pieces which are individually UV finite for plausible class of full propagators $S$. This is practically convenient, because for each group we can use the truncated version of full propagators without inducing any UV divergences.

Important applications in our mind are beyond mean field computations for dense quark equations of state, at densities relevant for the cores of neutron stars. There are suggestions that equations of state from hadronic to quark matter are crossover-like \cite{Baym:2017whm,Masuda:2012kf}, based on the observations of neutron star mass-radius relations as well as the microscopic understanding on states of QCD matter and relevant interactions \cite{Kojo:2014rca}. To discuss how hadronic matter continuously transforms into quark matter, we need a framework which enables us to describe not only single particle contributions from quarks and gluons, but also those from the hadronic objects as composite particles. The present work is a step toward such computations.

Another issue which lead us to the quark descriptions is the intricate structure in nuclear many-body interactions which are increasingly important beyond the empirical saturation point. The nuclear matter computations suggest that already at baryon density around twice nuclear saturation density, $n_B\sim 2n_0$, the contribution from each of 3-body force to the energy density is as large as 2-body force, but the sum of the former largely cancels, with which many-body effects appear to be smaller than the sum of 2-body contributions (e.g. see Table V in \cite{Akmal:1998cf}). Similar situations happen for hadronic models including $\sigma, \pi, \omega, \cdots$, etc., whose contributions show significant cancellations. We are not aware of why this should happen within nuclear computations, so it is desirable to look at microscopic descriptions with quarks to analyze the systematics of these cancellations.  These issues will be discussed elsewhere.

\section*{Acknowledgments}
The author acknowledges M. Alford, G. Baym, D. Blaschke, K. Fukushima, T. Hatsuda,  L. He, L. McLerran, R. Pisarski, and D. N. Voskresensky for comments and informing important references related to this work. The author also thanks U. Reinosa for pointing out issues relevant for this paper.
This work was supported by NSFC grant 11650110435 and a grant from the Simons Foundation. A part of this work was performed in part at Aspen Center for Physics (supported by National Science Foundation grant PHY-1607611), in JINR in Dubna, and completed during the stay at iTHEMS in RIKEN.

\section*{References}


\end{document}